\documentclass[12pt,preprint]{aastex}

\slugcomment{Article for DM 2004}

\shorttitle{Baryonic Dark Matter}
\shortauthors{ Schild}

\begin{document}


\title{The Detection and Nature of the Baryonic Dark Matter}


\author{Rudolph E. Schild}
\affil{Center for Astrophysics,
        60 Garden Street, Cambridge, MA 02138}
\email{rschild@cfa.harvard.edu}



\begin{abstract} Since the initial baryonic dark matter detection from
quasar microlensing was first announced in 1996, substantial strides have
been made in confirming the rapid microlensing signature in the Q0957
system and in other gravitational lens systems. The most rapid event
recognized had a 1 \% amplitude and a 12-hour duration.

Interpretation of the rapid fluctuations has centered upon three offered
explanations; microlensing of fine quasar structure by a population of
planet mass astronomical bodies in the lens galaxy, orbiting structures in
the accretion disc of the supermassive black hole of the quasar, or dark
clouds swarming around the luminous quasar source. The observations,
particularly the equal positive and negative fluctuations, seem to strongly
favor the cosmological population of planetary mass objects in the lens
galaxy.

Of the several ideas advanced for the origin of such a population, the most
appealing seems to be their birth at time of recombination 300,000 years
after the Big Bang.
\end{abstract}




\section{Introduction and Microlensing Results in Q0957}

At the 1997 Sheffield Dark Matter Conference I presented the conclusion
that the baryonic dark matter appeared to be a population of planet mass
bodies freely roaming in interstellar space of the lens galaxy at redshift
z=0.37 (Schild, 1996). This report was received with due skepticism;
``that's nice, but it's just Rudy's data for Rudy's quasar.'' Since that
time, the data have been confirmed from re-reduction of the original data
frames (Ovaldsen, 2003; Colley and Schild 1999, 2000). These authors have
confirmed the brightness estimates and the error estimates, and produced
evidence of a microlensing event of 12-hour duration and 1\% amplitude
(Colley and Schild 2003).

The Q0957 rapid microlensing confirmation was slow in coming because of a
conundrum in the detection; in order to detect rapid microlensing on 1-day
time scales, the A - B image arrival time delay had to be determined to a
tenth of a day, but such an accurate time delay would be difficult to
determine because of the rapid microlensing.

This conundrum was broken in the QuOC-Around-The-Clock project wherein a
10-night campaign in January 2000, followed by a repeat campaign in March
2002 (417 days later), with observatories around the globe, maintained
constant surveillance of the quasar lens brightness. Thus as the quasar
set at Mt. Hopkins in Arizona, it was rising in Hawaii, then in S. Korea,
Uzbekistan, Canary Islands, Montreal, and then again in Arizona.

The results of this project were published in Colley \& Schild et al
(2002, 2003), and produced the time delay 417.09 +/- .07 days, with some
discussion about the non-uniqueness of such estimates because the black
hole is a light emitting surface with 6 light hour size that is evidently
microlensed. The illuminated inner edge of the accretion disc would be a
factor of six larger, and outer luminous structure is also evident
(Vakulik \& Schild, 2003).

With this time delay, the authors were able to go back to observations
published previously and demonstrate convincing evidence for a microlensing
event of 12 hour duration and 1\% amplitude (Colley \& Schild, 2003, Figs 1
and 2). Cruder data sets published previously were unable to find such
elusive events (Schmidt and Wambsganss, 2000; Alcalde et al 2002).

It is important to put into context the tremendous advance in structure
resolution this represents. With overall quasar lensing produced by the
$10^{12} M_{\odot}$ mass of the lens galaxy, we produce lens image
separations of several arcsec. With milli-lensing, by globular cluster
mass bodies, we resolve structure on the milli-arcsec level. Microlensing
by individual stars in the lens galaxy resolves the outer structure of a
quasar, and nano-lensing by the planetary mass baryonic dark matter in the
lens galaxy resolves the internal structure of the quasar at the inner
region surrounding the supermassive black hole. At the next DARK MATTER
symposium, I expect to report the surface brightness of the black hole
event horizon from direct measurement.

\section{Microlensing Confirmation in Other Gravitational Lens Systems}

During the years when the Q0957 rapid microlensing was being confirmed,
additional lens systems were monitored and time delays found, and in all
cases, insofar as the brightness monitoring produced adequate data, the
microlensing fluctuations were found (Burud et al, 2000, 2001, 2002a,
2002b).  The amplitudes and durations found were comparable to those
already seen in Q0957. Perhaps the best example is the most recent,
HE1104-1805, where Schechter et al (2003) and Ofek and Maoz (2003) found
continuous microlensing with typical brightness amplitude of 10\% and
duration of 60 days. An important conclusion was in agreement with the
Q0957 discovery; equal positive and negative brightness events were found.

\section{ Alternative Explanations of the Data}

A fundamental prediction of the gravitational lensing theory is that for
microlensing in quasar lens systems, the optical depth to microlensing
should be approximately one, and so equal positive and negative events
should be seen. This is because the macrolensing producing the two quasar
images with arcsec separation produces a strong shear on large angular
scales, and the smaller microlensing events will enhance or diminish the
shear with about equal probability. So the signature of a cosmologically
significant microlensing population should be the equal positive and
negative events, as found by Schild (1999) from wavelet analysis in the
Q0957 system, and convincingly confirmed in HE1104 (Ofek and Maoz, 2003,
Figs. 4, 5).

Fortunately, other explanations have been explored. In a report that very
beautifully outlines the general problem, Gould and Miralde-Escude (1997)
consider the possibility that the fluctuations are produced by bright
point-like sources circulating in the quasar accretion disc and producing
brightness spikes when they cross the cusp pattern produced by stars in
the lens galaxy G1 of the Q0957 system. This scheme has been further
explored for HE1104 by Schechter et al (2003).

These schemes are frustrated by their failure to match the observed
properties: 1. They produce highly periodic effects not observed. 2. They
do not produce the equal positive and negative events observed
    and intrinsic to the unit optical depth microlensing case. 3. They
require super-relativistic speeds for the bright points, especially
    for the event of 12 hour duration reported by Colley and Schild. 4. Or
the scheme requires thousands of hot spots to get equal positive and
    negative events, producing large statistical fluctuations not observed.
5. No calculation is given to make plausible the existence of such compact
    blobs, each having the luminosity of a galaxy and a size smaller than
the
    quasar's central black hole. 6. Since the equal positive and negative
fluctuations are seen in two lens
    systems, it is not reasonable to imagine balanced bright points and dark
    clouds.

Another explored possible explanation by Wyithe \& Loeb (2003) imagines
that the fluctuations are simply caused by dark clouds orbiting the
quasar, at about the distance where the emission lines form (to avoid
super-relativistic speeds). But the authors note from their simulations
that the model does not produce brightness records that look like the
observed ones. It also requires dark clouds moving at relativistic speeds,
and does not address the problem of why such clouds would not be shredded
by differential rotation. Nor is it reasonable to imagine cool dark clouds
in the presence of CIII, CIV, and NV.

\section{The Origin and Nature of the Baryonic Dark Matter Particles}

If we provisionally accept the simplest explanation of the microlensing
observations, we must ask what is the origin of the planetary mass
particles. In the original Schild (1996) report of their detection, they
were called ``rogue planets'' because contemporaneous sub-mm observations
of young pre-planetary disc masses gave estimates of approximately $0.1
M_\odot$, and it was easy to imagine that most of this mass escaped to
produce a vast population of free-roaming planet mass objects. In the
intervening years, the proto-planetary disc mass estimates have declined
and the explanation seems less tenable.

Another possible explanation advanced by Carr and students (Barrows and
Carr, 1996), of a vast population of primordial black holes, seems not to
have received further theoretical or observational support.

By far the most interesting explanation comes from the fluid mechanics
community, which questions the neglect of viscosity, diffusion and
turbulence in the generally accepted Jeans 1905 acoustical criterion for
self-gravitational instability (Gibson 1996, 2000).  It is concluded that
when the viscosity of the plasma universe undergoes its dramatic decrease
at the time of recombination, 300,000 years after the Big Bang, the entire
gas universe fragments at Jeans and viscous-gravitational scales to form
globular-star-cluster-mass clumps of planetary-mass ``primordial fog
particles".  The publication Gibson (1996) proposing these dark clumps of
dark planets as the baryonic dark matter appeared within a month of the
independent observational result and ``rogue planet'' interpretation
published by Schild (1996).

Other predictions of the hydrodynamic theory of gravitational structure
formation include the top-down fragmentation of plasma starting 30,000
years after the Big Bang to form proto-superclusters, proto-clusters, and
proto-galaxies.  The massive weakly-collisional non-baryonic dark matter
diffuses to fill the voids between these structures, and fragments after
recombination to form large outer-galaxy and outer-galaxy-cluster halos.
Star formation occurs by accretion of the dark-matter planets (30 million
per star), leaving Oort Cloud size holes in an interstellar medium filled
with planets.  Evaporation of planets bordering the holes by
symbiotic-star (white dwarf and companion) plasma jets prior to SuperNova
Ia events provide a source of random systematic SNIa dimming that may be
an alternative to the ``dark energy'' interpretation.  Stars and luminous
galaxies with large central black holes form rapidly that can explain
observations of large red shift quasars and galaxies.  Hydro-gravitational
theory also explains the missing galaxy fragments expected in hierarchical
clustering scenarios.

\section{Some Further Consequences of the Baryonic Dark matter Detection}

Usually in science, when a major shift in understanding is made, a few
misinterpretations of previous data become evident. We summarize a few
such apparent shifts in prevalent paradigms now noticed.

The cometary knots in the Helix nebula illustrated and described in O'Dell
and Handron (1996) and previous references have been traditionally
explained as
Rayleigh-Taylor instabilities in the expanding gas shell, or as shock front
effects.  These explanations are untenable in view of the mass estimates
of Meaburn et al, (1998) which demonstrate mass enhancements
greater than 1000 over
the nebular gas density; these are untenable in the accepted models. We
believe that these are instead the primordial dark matter particles
of the type seen in
quasar microlensing and predicted by hydrodynamics (and called primordial fog
particles, PFP's). In this picture, the particles are at rest with respect
to the expanding gas shell, whereas in the other models they are expanding
with the gas shell. Proper motions can settle the matter, and the available
measurements by O'Dell et al (2002) give a measurement of the
expansion velocity of
only $4.5 \pm 9 km/sec$, where 14 km/sec is the known expansion velocity.
Thus measurements presently favor our predicted 0 km/sec, but do not yet
confirm our prediction.

The theory of solar system formation is stuck with a rocky planetary core
formation process; stones colliding with stones in pre-planetary discs
are believed to stick
together to form larger rocks, even though this process has never been
demonstrated in experiments. We believe that rocky cores of planets,
Kuiper belt objects, and Oort cloud objects were formed when the primordial
PFP population collected at its center the interstellar dust particles,
and then crushed them together when the PFP's froze as the universe cooled
below the 20 degree Kelvin freezing point of Hydrogen.

And we also note that our population of primordial particles would have an
important effect on the transparency of the universe to the light of
quasars and distant supernovae. If particles the sizes and masses of
the cometary knots in the Helix nebula are seen (without their tails)
to cosmological distance, they would intersect a significant fraction of
the light of these large distant objects, explaining the quasar peak at
z=1.9 and the dimming of distant supernovae. The complex processes of
absorption and refraction by PFP's need to be modeled to see whether the
grey extinction model for ``self replenishing dust'' (Goobar et al, 2002)
can be approximated
to reproduce the dimming of distant supernovae (Riess et al, 2004).


\begin{thebibliography}{}


\bibitem[Alcalde, D., et al, 2002]{alc02}
Alcalde, D., et al, 2002, \apj, 572, 729


\bibitem[Barrows ]{BC }
Barrows, J. \& Carr, B., 1996, Phys. Rev. D, 54, 3920

\bibitem[Burud et al 2000  ]{Bur00 }
Burud, I. et al, 2000, ApJ, 544, 117

\bibitem[Rurud et al 2001  ]{Bur01 }
Burud, I. et al, 2001, A\&A, 360, 805

\bibitem[Burud et al 2002a  ]{Bu02a }
Burud, I. et al, 2002a, A\&A, 383, 71

\bibitem[Burud et al 2002b  ]{Bu02b }
Burud, I. et al, 2002b, A\&A, 391,481

\bibitem[Colley \& Schild et al 2002  ]{CS02 }
Colley, W. N. \& Schild, R. et al, 2002, ApJ, 565, 105

\bibitem[Colley \& Schild  et al 2003  ]{CS03 }
Colley, W. N. \& Schild, R. et al, 2003, ApJ, 587, 71

\bibitem[Colley \& Schild 1999  ]{CS99 }
Colley, W. N. \& Schild, R., 1999, ApJ, 518, 153

\bibitem[Colley \& Schild 2000  ]{CS00 }
Colley, W. N. \& Schild, R., 2000, ApJ, 540, 104

\bibitem[Colley \& schild 2003  ]{CS03 }
Colley, W. N. \& Schild, R., 2003, ApJ, 594, 97

\bibitem[Gibson 1996  ]{GIB96 }
Gibson, C.H., 1996, Appl. Mech. Rev., 49,299; astro-ph/9904260

\bibitem[Gibson 2000  ]{Gib00 }
Gibson, C.H., 2000, J. Fluids Eng., 122, 830; astro-ph/0003352

\bibitem[Goobar et al 2002  ]{GOOB02 }
Goobar, A., Bergstrom, L., \& Mortsell, E., 2002, A\&A, 384, 1

\bibitem[Gould 1997  ]{G&J97 }
Gould, A. \& Escude-Miralde, J., 1997, ApJ, 483, L13

\bibitem[Meaburn et al 1998  ]{Mea98 }
Meaburn, J. et al 1998, MNRAS, 294, 201

\bibitem[O'Dell \& Handron 1996  ]{OH96 }
O'Dell, C.R., \& Handron, K. 1996, AJ, 111, 1630

\bibitem[O'Dell et al 2002  ]{OD02 }
O'Dell, C.R., et al, 2002, AJ, 123, 3329

\bibitem[Ofek \& Maoz 2003  ]{OM03 }
Ofek, E. \& Maoz, D., 2003, ApJ, 594, 101

\bibitem[Ovaldsen et al 2003  ]{OV03 }
Ovaldsen, J.E., et al 2003, A\&A, 402, 891

\bibitem[Riess et al 2004  ]{RIE04 }
Riess, A. et al 2004, astro-ph/0402512

\bibitem[Schechter et al 2003  ]{Sch03 }
Schechter, P. et al, 2003, ApJ, 584, 657

\bibitem[Schild 1996)  ]{Sch96 }
Schild, R., 1996, ApJ, 464, 125

\bibitem[Schild \& Vakulik 2003  ]{SV03 }
Schild, R., \& Vakulik, V., 2003, AJ, 126, 689

\bibitem[Schmidt \& Wambsganss 1998  ]{SW98 }
Schmidt, R., \& Wambsganss, J., 1998, A\&A, 335, 379

\bibitem[Schneider et al 1992  ]{SEE92 } 
Schneider, P., Ehlers, J. \& Falco, E., 1992,
``Gravitational Lenses'' [New York: Springer Verlag] p. 343

\bibitem[Wyithe et al 2003  ]{Wy03 }
Wyithe, J. \& Loeb, A., 2003, ApJ, 577, 615

\end{thebibliography}
\end{document}